\documentstyle[aaspptwo,11pt]{article}

\begin{document}

\centerline{\bf High Velocity Oblique Cloud Collision and}

\centerline{\bf Star and Star Cluster Formation through Gravitational
Instability}

\centerline{\bf of the Shock-Compressed Slab with Rotation and Velocity
Shear}

\bigskip
\bigskip

\centerline{Masatoshi U{\scriptsize SAMI},
Tomoyuki H{\scriptsize ANAWA}, and Mitsuaki F{\scriptsize UJIMOTO}}

\bigskip

\centerline{Department of Physics, School of Science, Nagoya University
Chikusa-ku, Nagoya
464-01}

\centerline{\it E-mail(MU) usami@a.phys.nagoya-u.ac.jp}

\vfill

\begin{abstract}

We study the gravitational instability of an isothermal
gaseous slab formed by cloud-cloud collision and compression at the
cloud interface.  The compressed gaseous slab rotates and has velocity
shear except when the collision is not exactly head-on.  The effects of
the rotation and velocity shear on the gravitational instability
have been evaluated for the first time.
We obtained the growth rate of perturbations as a function of the wavelength
for various gaseous slab models.  We also obtained the changes
in the density and velocity due to the perturbations.
Two types of unstable modes are found; one is due to the self-gravity
of the gaseous slab and the other is due to the velocity shear.
The former instability leads to the formation of star and star clusters.
Velocity shear decreases the growth rate of the former instability
while it increases that of the latter instability.
Rotation also decreases the growth of the former instability.
As the slab grows in mass, the growth rate of the self-gravitational
instability increases and eventually becomes greater than the
growing timescale of the slab.  Then a gravitationally bound condensation
is formed and its mass is larger when the slab has stronger velocity shear
and rotation.

\end{abstract}


\vfill

\eject

\section{Introduction}

The high-velocity cloud-cloud collision has been thought to be one of
the important triggering mechanisms of the star and star cluster formation.
When two clouds collide with supersonic velocities, dense gaseous
slab bounded by shock waves forms at the interface of
the colliding clouds.  The dense gaseous slab grows in mass and
becomes unstable against fragmentation because of the self-gravity.
The fragments collapse and evolve into stars and star clusters.

Many evidences have been found for star formation triggered
by the cloud-cloud collision.
NGC1333 (Loren 1976), G110-13
(Odenwald et al. 1992), W49N (Serabyn et al. 1993),
and Sagittarius B2 (Hasegawa et al. 1993) are nominated as the
present sites of massive star formation triggered by cloud collision.
The number density of Galactic H II regions is proportional to the
square of the local gas density, $ \rho _{\rm H_2} $.
This static suggests that substantial fraction of OB stars exciting the
H II regions  are formed by cloud-cloud collision (Scoville et al. 1986).

It is also proposed that the populous and globular star clusters are formed
by high-velocity gas clouds collisions. Fujimoto and Noguchi (1990),
Fujimoto and Kumai (1991a,b), Ashman and Zepf (1992), and
Kumai, Basu and Fujimoto (1993) assessed the gas-dynamical circumstances of
the Magellanic Clouds, M31, M33, M82, and other galaxies having
young populous
and globular star clusters, and found that these galaxies have large amount
of gas in large-scale unorganized motion induced internally or externally.
In a great contrast to these galaxies, our Galaxy has neither young globular
clusters, nor large-scale unorganized motion of gas.
{}From this comparison, they developed the model
that unorganized motion with velocity greater than
$\sim 50 {\rm km} \, {\rm s}^{-1}$ creates
strongly-compressed regions of gas through cloud-cloud collision,
and trigger the star cluster formation (see, Gunn 1980, Sabano and Tosa 1985,
and Kang et al. 1990 for another model of star cluster formation by
cloud-cloud collision).  In this model, the gas in the compressed slab
fragments into clumps whose mass is comparable to that of the whole star
cluster, and they cascade to smaller and smaller fragments, finally
they form a gravitationally bound stellar systems, - globular clusters.

In the star and star cluster formation model by cloud collision
one must take account of oblique collisions since they are much
more frequent than head-on collisions.  When two clouds collide
obliquely, the shock compressed gas slab rotates and is stressed
by the velocity difference in the tangential component (velocity shear).
The rotation and velocity shear may affect the fragmentation of
the shock compressed gas slab.  The present paper evaluates the
effects on the gravitational instability of the slab for the first
time.  It is shown that velocity shear increases the effective
sound speed of the gas and accordingly the minimum mass for
instability (the Jeans mass). Rotation also increases the minimum mass
for the instability.

In section 2 we describe our slab-model which mimics the
shock compressed gas formed by oblique collision of clouds.
Our slab takes account of rotation of the slab and velocity
shear in the slab.
In section 3, we formulate a linear analysis of perturbation
around the equilibrium state.  The numerical results are shown in
section 4 consisting of subsections 4.1 through 4.4 each of which
are for non-rotating slab without shear, non-rotating slab with
shear, rotating slab without shear, and rotating slab with shear,
respectively.
In section 5 we estimate the Jeans mass to apply the dispersion
relation obtained in the previous section to star and star cluster
formation.  Section 6 is devoted to discussion and conclusion.

\section{Dynamical Models for a Shock-Com-pressed Slab
    of Self-Gravitating Gas}

In the supersonic cloud-cloud collision, a shock-compressed gaseous slab is
formed at the interface of the two colliding clouds.
(e.g., Stone 1970; Struck-Marcell 1982; Sabano and Tosa 1985).
Figure 1 is the snapshot of a model of the oblique collision, with
the Cartesian coordinates referred to the gaseous slab, where
$x-y$ plane is taken on the interface and the $z$-axis is taken normal to it.
The gaseous slab rotates around the $y$-axis with angular velocity
$\Omega$, and the compressed gas flows parallel to the $x$ axis
with velocities $u_0$ directed toward $x = + \infty$ in the space
$z > 0$ and $x = - \infty$ in $z < 0$.  The shear flow is assumed
to be uniform in the $ x $- and $ y $- directions for simplicity,
$ u _0 \, = \, u _0 ( z ) $.
As shown in (c) of figure 1, the flow parallel to the $x$-axis
is dominant over that to the $z$-axis, because the latter component
is much reduced by the strong compression. We neglect the latter component
in the present paper.

\begin{figure}
\vspace*{15cm}
\caption{The obliquely colliding clouds. a: Snapshot of the ongoing collision.
Thick arrows are a reference frame. The clouds move along horizontal
lines and collide. We find a shock compressed slab at the interface of
two clouds. b: A model of the shock-compressed slab (dark patch).
Circles spread by a light patch denote the projected clouds on the $x-z$
plane. c: Gas flows in and out of the slab. The pre-shocked gas streams along
the horizontal lines, and compressed by the oblique shock.
Although the post-shocked gas flows along dashed arrows, we approximate the
stream line by solid arrows parallel to the slab.}
\end{figure}%

In order to simplify the problem, we take the following
three approximations.
First we neglect the high-temperature region immediately
behind the shock wave and assume the whole shock-compressed slab to
be isothermal.  The high temperature region is small in mass because
of fast radiative cooling therefrom.
Second the gaseous slab is assumed to be
uniform in the $x$- and $y$-directions.
As noticed by Stone (1970), the loss of gas across the lateral edges
is negligibly small during the period of the collision.
Third the gaseous slab is assumed to be in equilibrium,
although its column mass actually increases with time.
This equilibrium state can be achieved,
when the sound wave propagates from the shock front to
the interface before the gaseous slab grows considerably.
As Welter (1982) showed, it is actually realized in the supersonic
collision of gas clouds. When half of the slab thickness and isothermal sound
speed are denoted with $L$ and $c$ respectively, we have
for the sound crossing time $t_s$ between
$0 \leq z \leq L$,
$$ t_s \, \simeq \, { L \over c}.
\eqno(2.1)$$
Similarly the increasing time $t_m$ of the slab mass is written as
$$t_m  \, \simeq \, \sigma /\left( d \sigma \over dt \right)
\, \simeq \, {\rho_1 L \over \rho_0 c M_{\perp}}
\simeq \, M_{\perp} t_{\rm s} \, , \eqno (2.2) $$
where $\sigma$, $\rho_0$, $\rho_1$, and $M_{\perp}$ denote respectively
the column mass of the slab, preshock density,
postshock density, and the
Mach number of the gas incident on the shock front normally. Here we have
used relations,
$\sigma \sim 2 \rho_1 L$, $d \sigma / dt \sim 2 \rho_0 c M_{\perp}$,
and $\rho_1 / \rho_0 = {M_{\perp}}^2$.
We have thus quasi-equilibrium when $t_{\rm s} < t_{\rm m}$, and
$ M_{\perp} > 1,$ both of which hold in
the present supersonic collision of gas clouds.

\section{Normal Mode Analysis for a Plane-Parallel
Compressed Gaseous Slab
in Rotation and \ Nonuniform \ Parallel Streaming}

We analyze linear stability of a self-gravitating gaseous
slab with the shear along the $x$-axis and the rigid rotation
around the $y$-axis (figure 1). The basic equations governing the motion
of gas are written as,
$${\partial \mbox{\boldmath$u$} \over \partial t} +(\mbox{\boldmath$u$}
\cdot \nabla) \mbox{\boldmath$u$}
 + 2 \mbox{\boldmath$\Omega$} \times \mbox{\boldmath$u$} +
\mbox{\boldmath$\Omega$} \times (\mbox{\boldmath$\Omega$}
 \times \mbox{\boldmath$r$}) =$$
$$\hskip 4.0 true cm - \nabla \psi - {\nabla p \over \rho}, \eqno(3.1)$$
$${\partial \rho \over \partial t} + \nabla \cdot ( \rho \mbox{\boldmath$u$})
= 0 ,
\eqno(3.2)$$
and
$$ \triangle \psi =4 \pi G \rho,
\eqno(3.3)$$
where $\mbox{\boldmath$u$},\mbox{\boldmath$\Omega$},\mbox{\boldmath$r$},
\psi,p,\rho$ are
usual symbols, representing
the velocity of gas, angular velocity of the slab,
position vector, gravitational potential, pressure, and gas density,
respectively. As noted in section 2 and shown in figure 1,
the angular velocity of the slab is parallel to the $y$-axis,
and we have explicitly,
$$ \mbox{\boldmath$\Omega$} =(0,\Omega,0).
\eqno(3.4)$$
Since we assume an isothermal gas, the pressure is written as
$$p = {c}^2 \rho .
\eqno(3.5)$$
The dynamical quantities $\mbox{\boldmath$u$}, \rho , \psi , p$
are expressed as the sum of those in an equilibrium state and
the deviations from it, the former and latter of which are
specified with suffix 0 and with hat and suffix 1, respectively,
like for example,
$$ \rho (\mbox{\boldmath$r$},t) = \rho _0 (z) +{\hat \rho}_1
(\mbox{\boldmath$r$},t)\; .
 \eqno(3.6)$$
The velocity in equilibrium, $\mbox{\boldmath$u$}_{0}$, has only the
$ x $-component and is a function of $z$,
$$\mbox{\boldmath$u$}_0 = \lbrack u_{0}(z), \, 0, \, 0 \rbrack \; .
\eqno(3.7)$$
Note that the other equilibrium quantities $ p_{0} $ and
$ \psi _{0}$
are also functions of $ z $ and independent of
$t$, $x$ and $y$, since we have assumed that the slab is homogeneous
in the $x$- and $y$-directions.

\subsection{Equilibrium \ Configuration of the Rotating Gaseous Slab
with Shear}

Substituting equation (3.6) and similar equations for other variables,
into equations (3.1), and (3.3), and using equations (3.4), (3.5),
and (3.7),
we derive the equations for the equilibrium state, by omitting the
terms with hatted symbols,
$$ 2 \Omega u_0 = \rho_0 { d \psi_0 \over dz} +{c}^2 {d \rho_0 \over dz},
 \eqno(3.8)$$
and
$${d^2 \psi_0 \over dz^2} = 4 \pi G \rho_0.
\eqno(3.9)$$
Equation of mass conservation (3.2) is automatically satisfied.
Equation (3.8) represents that the Coriolis force per unit mass
balances the gravitational acceleration and pressure gradient.
Three variables, $u_0, \rho_0, $ and $ \psi_0$ are constrained by
only two equations (3.8) and (3.9),
and hence we have infinitely large number of solutions.
In this paper we restrict ourselves to the case that the density
distribution has the same functional form to that given by Spitzer (1942),
$$\rho_0 (z) = \rho_c {\rm sech}^2 \left( z \over H_R \right).
\eqno(3.10)$$
Then we obtain,
$$u_0 (z) = U {H_R \over H}
{\rm tanh} \left(z \over H_R \right),
\eqno(3.11)$$
$$\psi_0 (z) = 4 \pi G \rho_c {H_R}^2 \ln \left[ \cosh \left( z \over
H_R \right) \right],
\eqno(3.12)$$
with
$$ H_R \equiv { H \over \sqrt{ 1- \Omega U H /
 {c}^2 }} \eqno(3.13)$$
${\rm with} \; \; \; \;
   \Omega U H /c^{2}<1,$
and
$$ H \equiv { c \over \sqrt{2 \pi G \rho_c}},
\eqno(3.14)$$
where $\rho_c$ is the density at $ z=0 $, and $H_R$ is a density
scale height which tend to $H$ as $\Omega$ to zero.
The velocity shear is $ d u _0 / d z \, = \, U / H $
on the $ z \, = \, 0 $ plane.  The column density $ \sigma $ is
given by
$$ \sigma \, \equiv \, \int _{-L} ^{L} \rho \, dz \, = \,
2 \rho _c H _R \, {\rm tanh} \left( L \over H _R \right) \; . \eqno (3.15) $$

\subsection{Linear Perturbed Equations}

Linearized equations of equations (3.1) to (3.3)
and (3.5) are written as, by using equations (3.4), (3.6) and (3.7),
$$ {\partial {\hat u}_{x1} \over \partial t} + u_0 {\partial {\hat u}_{x1}
\over \partial x} +{\hat u}_{z1} {\partial u_0 \over \partial z}
+2 \Omega {\hat u}_{z1} = \hskip 1.5 true cm$$
$$\hskip 3.0 true cm -{\partial {\hat \psi}_1 \over \partial x}
-{\partial \over \partial x} \left( {\hat p}_1 \over \rho_0 \right),
\eqno(3.16)$$
$$ {\partial {\hat u}_{y1} \over \partial t} + u_0 {\partial {\hat u}_{y1}
\over \partial x} = -{\partial {\hat \psi}_1 \over \partial y} -
 {\partial \over \partial y} \left( {\hat p}_1 \over \rho_0 \right) ,
\eqno(3.17)$$
$${\partial {\hat u}_{z1} \over \partial t} +u_0 {\partial {\hat u}_{z1}
\over \partial x} -2 \Omega {\hat u}_{x1} = \hskip 2.0 true cm$$
$$\hskip 3.0 true cm -{\partial {\hat \psi}_1 \over
\partial z} -{\partial \over \partial z} \left( {\hat p}_1 \over \rho_0
\right) ,
\eqno(3.18)$$
$${\partial {\hat \rho}_1 \over \partial t} +u_0 {\partial {\hat \rho}_1
\over \partial x} +\rho_0 {\partial {\hat u}_{x1} \over \partial x} +
\rho_0 {\partial {\hat u}_{y1} \over \partial y} + \hskip 2.0 true cm$$
$$\hskip 3.5 true cm +{\partial \over
\partial z} \left( \rho_0 {\hat u}_{z1} \right) = 0 ,
\eqno(3.19)$$
$$\triangle {\hat \psi}_1 =4 \pi G {\hat \rho}_1 ,
\eqno(3.20)$$
$$ {\hat p}_1 ={c}^2 {\hat \rho}_1 .
\eqno(3.21)$$
Since the equilibrium quantities are functions only of $z$,
the perturbed ones can be expressed in the form of, for example,
  $$ {\hat \rho}_1 (\mbox{\boldmath$r$},t) = \rho_1 (z) {\rm exp}
  \lbrack i(k_x x +k_y y -\omega t) \rbrack \; .\eqno(3.22)$$
We introduce new variable $\mbox{\boldmath$y$} (z)$, consisting of
four component, $y_1$ through $y_4$
$$ \mbox{\boldmath$y$} (z) \; = \;
\left[ \matrix{y_1 \cr y_2 \cr y_3 \cr y_4 } \right] \equiv
\left[ \matrix{p_1 \cr {i \rho_0 u_{z1} / \omega}
\cr \rho_c \psi_1 \cr \rho_c g_{z1} } \right],
\eqno(3.23)$$
where $g_{z1}$ denotes the perturbed gravity in the $z$-direction,
$$ g_{z1} \equiv {d \psi_1 \over dz} \; .
\eqno(3.24)$$
After straightforward manipulation we have for equations (3.16) through
(3.21),
$${d \mbox{\boldmath$y$} \over dz} = \mbox{\boldmath$A$} \mbox{\boldmath$y$},
\eqno(3.25)$$
where the $4 \times 4$ matrix $\mbox{\boldmath$A$}$ has following elements
$$A_{11} = {1 \over \rho_0} { d \rho_0 \over dz} - {2 \Omega k_x \over
k_x u_0 - \omega},
\eqno(3.26)$$
$$A_{12} = - \omega ( k_x u_0 - \omega ) + \hskip 2.0 true cm$$
$$\hskip 2.5 true cm +{2 \Omega \omega \over k_x u_0 -
\omega} \left( { d u_0 \over dz} + 2 \Omega \right),
\eqno(3.27)$$
$$A_{13} = -{2 \Omega k_x \over k_x u_0 - \omega} {\rho_0 \over \rho_c},
\eqno(3.28)$$
$$A_{14} = -{\rho_0 \over \rho_c},
\eqno(3.29)$$
$$A_{21} = \left( {c}^2 \over \omega \right) \left( k_x u_0 - \omega \right)
- \hskip 2.0 true cm $$
$$\hskip 2.0 true cm -{1 \over k_x u_0 - \omega }
 \left( {k_x}^2 +{k_y}^2 \over \omega \right),
\eqno(3.30)$$
$$A_{22} = {1 \over k_x u_0 - \omega} \left[ {d \over dz} (k_x u_0) + 2 \Omega
k_x \right],
\eqno(3.31)$$
$$A_{23} = - {1 \over k_x u_0 - \omega} {{k_x}^2 + {k_y}^2 \over \omega}
 {\rho_0 \over \rho_c} ,
\eqno(3.32)$$
$$A_{34} = 1,
\eqno(3.33)$$
$$A_{41} = {4 \pi G \rho_c \over {c}^2 },
\eqno(3.34)$$
$$A_{43} = {k_x}^2 + {k_y}^2 ,
\eqno(3.35)$$
and others are zero, i.e., $A_{24} = A_{31} = A_{32} = A_{33} = A_{42}
= A_{44} = 0$.

We integrate equation (3.25) and seek solutions satisfying the boundary
conditions described in subsection 3.3.  Then we obtain the dispersion
relation, i.e., $ \omega $ as a function of $ k _x $ and $ k _y $
for a given slab model.

\subsection{Boundary Conditions}

Since the \ gaseous slab is \ sandwiched and pressed together with two
plane-parallel shock fronts, the boundary surface of our integration is
assumed to be rigid and fixed [e.g. Stone (1970), Voit (1988)].
That is, we take into account the fact that the plain shock front is stable
against perturbation (e.g. Landau and Lifsitz 1987).
When a wavy perturbation is
imposed along the shock front, the velocity component of gas
parallel to the disturbed surface does not change across the shock front,
but, the component normal to it decreases considerably due to
the compression of gas. The resultant motion of the gas is divergent
from the convex region (seen from the downstream) and similarly convergent to
the concave one, restoring the distorted front surface to its original
plane surface.

We allow of the pressure discontinuity $\delta p$ due to the perturbation.
According to equation (3.23), it is written as
    $$ y_1 (L) = \delta p.
    \eqno(3.36)$$
The perturbed velocity normal to the surface, $u_{z1}$ or $y_2$,
must be zero,
    $$ y_2 (L) = 0.
    \eqno(3.37)$$
The perturbed potential $\psi_1$ is continuous and smooth at $z= \pm L$.
Solving the Poisson's equation in the space outside the slab, $\triangle
{\hat \psi_1} = 0 $, we have
   $$\psi_1 (z) = \psi_1 (L) \exp[ - \sqrt{{k_x}^2 + {k_y}^2} (
 z - L)],
   \eqno(3.38)$$
in $z > L$, and thus
   $${d \over dz} \psi_1 (L)
     = - \sqrt{{k_x}^2 + {k_y}^2} \psi_1 (L).
     \eqno(3.39)$$
Using equations (3.23)
and (3.24), we have the boundary conditions on $y_3$ and $y_4$,
  $$y_3 (L) = \rho_c \psi_1 (L),
   \eqno(3.40)$$
and
   $$y_4 (L) = -\rho_c \sqrt{{k_x}^2 +
    {k_y}^2} \psi_1 (L) .
    \eqno(3.41)$$
Equations (3.36), (3.37), (3.40), and (3.41) form a complete set of
the boundary conditions at $z=L$.
Note, however, that it includes two parameters, $\delta p$ and $\psi_1 (L)$,
whose numerical choices are still free.

Similar boundary conditions are imposed at $ z \, = \, - L $.  In total
the boundary conditions have four degrees of freedom as well as the
differential equation (3.25) has.

\subsection{Numerical Methods}

Our method to obtain the numerical dispersion relation is
essentially the same as those of Nakamura et al. (1991) and
Matsumoto et al. (1994).  We seek $ \omega $ for which a solution
of equation (3.25) satisfys the boundary conditions at
$ z \, = \, \pm L $ using a bisection method.

The general solution satisfying the boundary condition at $ z \, = \, L $ is
expressed as a linear combination of two linearly independent
solutions,
$$ \mbox{\boldmath$y$} (z) = \delta p \mbox{\boldmath$y$}^{(1)} (z) +
\psi_1 (L) \mbox{\boldmath$y$}^{(2)} (z)
\eqno(3.42)$$
where $ \mbox{\boldmath$y$}^{(1)}(z) $ and $ \mbox{\boldmath$y$}^{(2)}
(z) $ are solutions of equation (3.25) and their values at
$ z \, = \, L $ are given by
$$ \mbox{\boldmath$y$}^{(1)} (L) \, = \,
\pmatrix{ 1 \cr 0 \cr 0 \cr 0 \cr} \, ,
\eqno(3.43)$$
and
$$ \mbox{\boldmath$y$}^{(2)} (L) \, = \,
 \pmatrix{ 0 \cr 0 \cr \rho_c  \cr - \rho_c \sqrt{ {k_x}^2 + {k_y}^2} \cr}
\, ,
\eqno(3.44)$$
respectively.  By the Runge-Kutta method we integrated equation
(3.25) to obtain $ \mbox{\boldmath$y$}^{(1)} (z) $ and
$ \mbox{\boldmath$y$}^{(2)} (z) $.  The parameters, $ \delta p $ and
$ \psi (L) $ are left unfixed at this moment.

Similarly \ we \ obtain the \ general solution, $ \mbox{\boldmath$y$} (z)
\, = \, \delta p ^\prime \mbox{\boldmath$y$}^{(3)} (z) \, + \,
\psi (-L) \mbox{\boldmath$y$} ^{(4)} (z) $, which satisfys
the boundary condition at $z = -L$.  The two general solutions
coincide each other at $ z \, = \, 0 $, and thus
$$ \pmatrix{ c_{11} & \ldots & c_{14} \cr
             \vdots & \ddots & \vdots \cr
             c_{41} & \ldots & c_{44} \cr}
\pmatrix{ \delta p \cr
          \psi _1 (L) \cr
          -\delta p' \cr
          -\psi _1 (-L) \cr} = {\bf 0},
\eqno(3.45)$$
with
$$ c_{ij} \equiv {y_i}^{(j)} (0) \, \, \, \, \, i,j = 1,2,3,4.
\eqno (3.46)$$
Equation (3.45) has a non-trivial solution when and only when
the determinant of the matrix $ ( c _{i,j} ) $ vanishes,
$$ {\rm det} \; \left\vert \matrix{ c_{11} & \ldots & c_{14} \cr
             \vdots & \ddots & \vdots \cr
             c_{41} & \ldots & c_{44} \cr} \right\vert = 0
\eqno(3.47)$$
We computed the determinant as a function of $ \omega $ and
sought $ \omega $ for which the determinant has a sufficiently small value.
See Matsumoto et al. (1994) for the algorithm to
obtain $ \omega $ for the determinant has a smaller value
successively.

\section{Gravitational Instability of the Plane-Parallel Gaseous Slab}

The dispersion relations between $\omega$ and $\mbox{\boldmath$k$}
[\equiv (k_x , k_y)]$
are obtained for four cases with/without the shear and the rotation.
First we study the gravitational instability of non-rotating slab
without velocity shear in subsection 4.1.  The effects of velocity
shear and rotation are shown separately in subsections 4.2 and 4.3,
respectively.  We discuss the gravitational instability
of a rotating slab with velocity shear in subsection 4.4.

We represent all the dispersion relation in a nondimensional form for extensive
applications of the results.  For this purpose we normalize velocity, time,
and density, with the isothermal sound speed $c$,
the free-fall time $1/\sqrt{2 \pi G \rho_c}$, and
the density $\rho_c$ at $z=0$, respectively. Thereby the length is
automatically normalized with the density scalehight, $ H \, = \,
c / \sqrt{2 \pi G \rho _c } $.
After the normalization, parameters are reduced to $\mbox{\boldmath$k$},
L, U,$ and $\Omega$.

\subsection{Non-rotating Slab without Velocity Shear ($\Omega = U = 0$)}

\begin{figure}
\vspace*{6cm}
\caption{Dispersion relations of the gravitational instability when
$U= \Omega = 0$. Thick and thin lines
are results for the fixed and free boundaries,
respectively. We write adopted $L$ in the figure.}
\end{figure}

Figure 2 shows the squared growth rates ($-\omega^2$) of the unstable
perturbation as functions of $k_x$ for $ k _y \, = \, 0 $.
The thin and thick curves denote the dispersion relations for
$ L \, = \, 0.5 $ and 1.0, respectively.  The growth rate
vanishes at $ k _x \, = \, 0 $ and the perturbation is stable
when the wavenumber exceeds the critical one, $ k _x \, > \, k _{\rm cr} $.
 The critical wavenumber is
$ k _{cr} \, = \, 0.74 $ and 0.95 for $ L \, = \, 0.5 $ and 1.0,
respectively.   The growth rate has its maximum $ - \omega _{\rm max} {}^2 $
= 0.17 and 0.34 at $ k _{x \, {\rm max}} \, = \, 0.37 $ and 0.46,
respectively for $ L \, = \, 0.5 $ and 1.0.

\begin{figure}
\vspace*{6cm}
\caption{The dependence of the maximum growth rates and their wavenumbers on
the thickness of the slab for $U = \Omega = k_y = 0$.
The meanings of the thick and thin lines are the same as in figure 2.
Upper two lines represents squared maximum growth rates, while lower ones are
the wavenumbers at the maximum growth rates.}
\end{figure}

Figure 2 suggests that $-{\omega_{\rm max}}^2$ and $k_{x \, {\rm max}}$
decrease as the slab thickness $2 L$ decreases.  Figure 3
confirms it.  For a given $ k _x $ the mass contained in a
wavelength decreases and hence the growth of the gravitational
instability weakens as the slab becomes thinner.
When the slab is thin, $0 \leq L
\leq 0.5$, the dispersion relation shown in figure 2 and 3 can be
reproduced approximately by
the gravitational instability of the infinitely thin sheet of gas whose
column density is given by equation (3.15) (e.g. Stone 1970, and
Goldreich and Tremaine 1979):
$$ \omega^2 = {c}^2 {k_x}^2 - 2 \pi G \sigma k_x,
\eqno(4.1)$$
with
$$-{\omega_{\rm max}}^2 = {(\pi G \sigma)^2 \over {c}^2},
\eqno(4.2)$$
and
$$k_{x \; {\rm max}} = {\pi G \sigma \over {c}^2}
\eqno(4.3)$$
[We suspend the normalized expression for quantities in equations
(4.1) through (4.3).].
Also for a larger $ L $ equations (4.2) and (4.3) reproduce the
qualitative features of the dispersion relation but less accurately.
At the limit of $ L \rightarrow \infty $,
$-{\omega_{\rm max}}^2$ and $k_{x \, {\rm max}}$ tend to $0.45$ and
$0.47$, respectively (See also Simon 1965), while equations (4.2)
and (4.3) give $ -{\omega_{\rm max}}^2 \, = \, 1.0 $ and
$ k_{x \, {\rm max}} \, = \, 1.0 $.

\begin{figure}
\vspace*{8cm}
\caption{The cross section of the slab perturbed by the fastest growing
perturbation, when $U = \Omega = k_y = 0$.}
\end{figure}

The top diagram of figure 4 represents a cross sectional view of the
fastest growing density perturbation with the growth rate,
$- {\omega_{\rm max}}^2 = 0.17$, for the slab of $L =0.5$.
It delineates the one wave-length region
$ - \pi / k_{ x \, {\rm max}} \leq x \leq \pi / k_{x \, {\rm max}}$.
In order to see the vertical structure, we magnify the
slab thickness about ten times in the bottom diagram.
Contours denote the gas density.
Arrows in the bottom panel shows the velocity distribution.
We find that the perturbed velocities are
predominantly parallel to the $x$-axis, and negligibly small
in the $z$-axis, that is,
the perturbation grows due mainly to the gas flow parallel to the slab.
This tendency is
more prominent when the slab is thinner.

\subsection{Non-Rotating \ Slab with \ Velocity Shear
($\Omega = 0, U \not= 0$)}

In the shearing slab both the gravitational and Kelvin-Helmholtz
modes appear. The dispersion relations described here are about the
gravitational mode, which continuously approach those in the previous
subsection 4.1, as the shear becomes weaker. (We will discuss briefly
the Kelvin-Helmholtz modes in subsection 6.2.)

\begin{figure}
\vspace*{6cm}
\caption{Dispersion relations when $U \not= 0$ and $\Omega = 0$.
The thin line denotes the relation for the non-shearing slab.}
\end{figure}

\begin{figure}
\vspace*{6cm}
\caption{The dependence of the squared maximum growth rate on the shear
magnitudes.
The thin line is ${\omega^2}_{\rm max} \propto {M_{\parallel}}^{-2}$.}
\end{figure}

Figure 5 shows the squared growth rate $-{\omega}^2$ as functions of
$k_x$, for three cases that $M_{\parallel} \equiv u_0 (L) / c = 0,
0.5,$ and $1.0$, when $L = 1.0$ and $k_y =0$. Both $-{\omega_{\rm max}}^2$
and $k_{x \, {\rm max}}$
decrease as $M_{\parallel}$ increases.
The dispersion relation remains parabolic-like on the $ \omega ^2 - k $
diagram.
Figure 6 shows the detailed
behaviour of $-{\omega_{\rm max}}^2$ against $M_{\parallel}$ for the
gaseous slabs of $L = 1.0, 0.5,$ and $0.3$, when $\Omega = 0$ and $k_y = 0$.
For a given $L$, $-{\omega_{\rm max}}^2$ is smaller when the
shear is stronger.
Particularly, when $M_{\parallel} \gg 1$, we have
$ -{\omega_{\rm max}}^2
\propto {M_{\parallel}}^{-2}$ (Compare with the thin line.).

According to our extensive calculation the
numerical dispersion relation can be well approximated by
$$ \omega^2 = {c}^2 (1 + {M_{\parallel}}^2) {k_x}^2
- 2 \pi G \sigma k_x .
\eqno(4.4)$$
Equation (4.4) implies that the velocity shear may be dealt with
as an ensemble of vortices and the gaseous slab has the effective
sound speed, $ c \, \sqrt{ 1 \, + \, M _\parallel {} ^2 } $
[See equation (4.1).].
Note that the square of the effective sound speed is the arithmetic
sum of the square of the sound speed and that of velocity shear.
{}From equation (4.4) we obtain
$$-{\omega_{\rm max}}^2 = {(\pi G \sigma)^2 \over {c}^2 (1 +
{M_{\parallel}}^2)},
\eqno(4.5)$$
and
$$k_{x \; {\rm max}} = {\pi G \sigma \over {c}^2 (1 + {M_{\parallel}}^2)}.
\eqno(4.6)$$
Equation (4.5) and (4.6) are also consistent with our numerical
results, e.g. $-{\omega_{\rm max}}^2 \propto {M_{\parallel}}^{-2}$
when $M_{\parallel} \gg 1$.

\begin{figure}
\vspace*{8cm}
\caption{The cross section of the shearing slab perturbed by the fastest
growing perturbation.}
\end{figure}

Figure 7 is the same as figure 4 except that the shear is present.
$k_x = 0.285, k_y = 0.0, L = 1.0,$ and $M_{\parallel} = 1.0$ are applied.
The concentration of gas is similar to that in figure 4, but
the perturbed gas spirals-in toward the point of maximum density.
Note, however, that the spirals are elongated along the $z$-axis.

When $ \Omega \, = \, 0 $, the wave number, $ ( k _x , \, k _y ) $
appears in the matrix, $ \mbox{\boldmath$A$} $, and in the boundary
conditions only in the form of
$ k \, = \, \sqrt{ k _x {} ^2 \, + \, k _y {} ^2 } $ and $ k _x u _0 (z) $.
Thus, we obtain a relation,
$$ \omega ( k _x , \, k _y ; \, u _0 , \, \Omega = 0 ) \; = \hskip 4.0 true
cm$$
$$\omega ( {k _x}^\prime = \sqrt{ {k _x} ^2 + {k _y} ^2 } ,
{k _y }^\prime = 0 ; k _x u _0 / {k _x} ^\prime ,
\Omega = 0 ) \; . \eqno(4.7)$$
Using this relation we can obtain the growth rate for a mode having
non-vanishing $ k _y $. Hence, the discussions about the dispersion
relation (figure 5) and the correlation between the maximum growth rate
and the Mach number of the streaming motion parallel to the slab (figure 6)
remain unchanged after the transformation from $k_x$ and $M_{\parallel}$ to
$k $ and $M_{\parallel} k _x / k $.  The effective sound speed
can be expressed as $ c _{\rm eff} {} ^2 \, = \, c ^2 \,
\lbrack 1 \, + \, k _x {} ^2 M _\parallel {} ^2 / ( k _x {} ^2 \, + \,
k _y {} ^2 ) \rbrack $.  When $ k _x \, = \, 0 $ and
$ k _y \, \neq \, 0 $, the growth rate is independent of the shear strength.

\subsection{Rotating Slab without Shear
($\Omega \not= 0, U = 0$)}
\begin{figure}
\vspace*{6cm}
\caption{Dispersion relations when $\Omega \not= 0$ and $U = 0$.
The thin line corresponds to the non-rotating slab.}
\end{figure}

Figure 8 shows the squared growth rates $-{\omega}^2$ as a function of $k_x$
for $\Omega = 0$, $ 0.5 $, and $1.0$.  The parameters $ L $ and $ k _y $
are kept constant, $L = 1.0$ and $k_y =0$. The growth rate
decreases monotonically, as $\Omega$ increases.
The rotation tends to suppress the gravitational instability.
However, the decrease in the growth rate is smaller than that for
a rotating disk of which rotation axis is the normal to the disk.
The latter growth rate is approximated by
$$ \omega^2 = {c}^2 {k_x}^2 - 2 \pi G \sigma k_x + 4 {\Omega}^2 \ ,
\eqno(4.8)$$
(see, e.g. Goldreich and Tremaine 1979) and the growth rate
decreases by $ 4 \Omega ^2 $ at a given $ k _x $.  Accordingly,
as $ \Omega $ increases, $ k _{x \, {\rm cr}} $
decreases while $ k _{x \, {\rm max}} $ remains unchanged.
On the other hand, $ k _{x \, {\rm max}} $ decreases in our model
as $ \Omega $ increases.  The critical wavenumber $ k _{x \, {\rm cr}} $
is independent of $ \Omega $ in our model.  As proved in  appendix,
the velocity perturbation vanishes for the marginally stable mode
of $ k _x \, = \, k _{x \, {\rm cr}} $ in our model.  Then the
Coriolis force does not work in the limit of $ i \omega \, \rightarrow
\, + 0 $.  As $ \Omega $ increases, the growth rate decreases but
never vanishes as far as the mode is unstable at $ \Omega \, = \, 0 $.

\begin{figure}
\vspace*{6cm}
\caption{The dependence of the squared maximum growth rate on
the angular velocity.}
\end{figure}

The correlation between maximum growth rate $-{\omega_{\rm max}}^2$
and the angular velocity, $\Omega$, of the slab is given
in figure 9 for $L = 0.1, 0.3, 0.5,$ and $1.0$, when $k_y =0$ and $U =0$.
As $ \Omega $ increases, $-{\omega_{\rm max}}^2$ decreases but a little
and the decrease is very small for $L \leq 0.1$.

These new results are intrinsic to the rotating slab sandwiched and pressed
together with two plane-parallel rigid surfaces; when the gaseous component
is perturbed, it moves easily along the slab, but the Coriolis force due to
this motion drives the gas to move normal to the slab. The narrowly-separated
rigid surfaces, however, suppress this motion. The suppression is extremely
strong, when the slab is thin, or the wavelength is long. Thus, the dispersion
relations lack dramatic effect of the rotation in that case.

\begin{figure}
\vspace*{17cm}
\caption{The cross section of the rotating slab disturbed by the fastest
growing perturbation. The velocities seen from the rotating frame
are delineated. We assume that $\Omega = 0.4$ and $k_x = 0.43$ in figure (a),
and $\Omega = 1.0$ and $k_x =0.30$ in figure (b).}
\end{figure}

Figure 10a shows the perturbed density distribution and is
the same as figure 4 except for $\Omega = 0.4$.
We have taken a case of moderate slab thickness: $L = 1.0$,
and assumed that $k_y = 0$.
Then, the maximum growth rate is
$-{\omega_{\rm max}}^2 = 0.25$ and the corresponding wave number
is $k_{x \, {\rm max}} = 0.43$. The global features are similar to those
in figure 4, but we find that the iso-density contours are
tilted due to the Coriolis force. The velocity normal to the slab
is actually seen, derived by the Coriolis force.
The concentration of gas is due primarily to the $x$-component
of the perturbed velocity.

As the angular velocity increases the iso-density contours become
more elongated and tilted, and then form two separated concentrations
of gas (figure 10b).

So far we have dealt with the case of $ k_x \not= 0$ and $k_y = 0$.
When $k_y \not= 0$ and $k_x = 0$, or the wave vector is parallel
to the rotation axis, we find the growth rates hardly change due
to the rotation, since the perturbed flow is nearly parallel to the
rotation axis, and then $2 \mbox{\boldmath$\Omega$} \times
\mbox{\boldmath$u$}_1 \simeq 0$.

\subsection{Rotating Slab with Shear
($\Omega \not= 0, U \not= 0$)}

This case would be the most realistic gaseous slab occurring at
the interface of the obliquely-colliding clouds with supersonic velocity.
As seen from equation (3.13), the equilibrium state is valid
only in the parameter range of $\Omega U H /c^2 < 1$.
Therefore, we present the results within it.

\begin{figure}
\vspace*{6cm}
\caption{The dispersion relations when $U \not= 0$ and $\Omega \not= 0$.
Adopted $M_{\parallel}$ and $\Omega$ for each line are written in the figure.}
\end{figure}

Figure 11 shows the dispersion relation between $k_x$ and $-{\omega}^2$ for
$M_{\parallel} = 1.0$, $\Omega = 0.5$, $k_y = 0.0 $, and $L = 1.0$.
For comparison, the dispersion relations for
$ ( M_\parallel , \, \Omega ) \, = \, (0.0, \, 0.0) $, $ (1.0, \, 0.0 ) $,
and $ (0.0, \, 0.5) $ are shown with thin curves for comparison.
The wavenumber and the thickness of the slab are kept constant
$ k _y \, = \, 0 $ and $ L \, = \, 1.0 $ for all the thin and thick
curves.  The gravitational instability is suppressed additively
by shear and rotation, and the critical wave number $k_{x \, {\rm cr}}$
decreases as $\Omega$ increases, which is derived to be constant when $U = 0$.

\begin{figure}
\vspace*{6cm}
\caption{The dependence of the squared maximum growth rate on the angular
velocity.}
\end{figure}

Figure 12 shows the squared maximum growth rates as functions of
$\Omega$ for $k_y = 0$ and $L = 1.0$. Three curves represent
$ - \omega _{\rm max} {} ^2 $ for $M_{\parallel} = 0, 0.5,$ and $1.0$.
In this figure we confirm again that the rotation and shear suppress the
gravitational instability.  Note that the maximum growth rate is
larger for a larger $ M _\parallel $ in the region of $ \Omega \, \ga
\, 1.1 $.  This is because the column density $\sigma$ is larger for a larger
$ M _\parallel $ for given $ L $ [See equation (3.15).].
A larger column density brings a larger maximum growth rate.

\begin{figure}
\vspace*{8cm}
\caption{The cross section of the shearing and rotating slab disturbed
by the fastest growing perturbation.}
\end{figure}

Figure 13 is similar to both of figures 7 and 10a, but for
$M_{\parallel} = 1.0,$ $\Omega = 0.5,$, $L = 1.0$,
$k_x =0.2$, and $k_y = 0$.
The contour curve of constant density slants to the $ x $-direction
as in figure 10 a and the gas stream spirals into the point of the
maximum density as in figure 7.

The dispersion relation between $k_x$ and $-{\omega}^2$ in figure 12
cannot be empirically represented in a simple form including the parameters,
 the slab thickness $2 L$, shear U, and angular velocity $\Omega$.
However, when the slab is a thin sheet of gas and the Coriolis
force more suppressed, the dispersion relation reduced to that of
the non-rotating slab in equation (4.4).

\section{Jeans Mass with Rotation and Shear}

When the gravitational instability grows and non-linear contraction
proceeds, a high-density gaseous clump forms in the slab. When the slab
rotates and has velocity shear, the Jeans wave length depends on the direction
of the wave vector: The one parallel to the $x$-axis is given by
$\lambda_{x \, {\rm J}} = 2 \pi / k_{x \, {\rm max}}$ and equation (4.6),
when $L \ll H_{\rm R}$, and the other one, $\lambda_{y \, {\rm J}}$,
parallel to the $y$-axis remain unaffected by the shear and rotation,
and obtained from equation (4.3).
Therefore, the most probable Jeans mass $M_{\rm J}$ would be
$$ M_{\rm J} = {\lambda_{x \, {\rm J}} \over 2}{\lambda_{y \, {\rm J}} \over
2} \sigma,
\eqno(5.1)$$
where $\lambda_{y \, {\rm J}} \equiv 2 \pi / k_{y \, {\rm max}}$.
Note that $\lambda_{x \, {\rm max}}, \lambda_{y \, {\rm max}}$, and $\sigma$
are functions of the slab thickness $2L$, which increases with time.
Using a similar
equation, Whitworth et al. (1994) calculated the mass
fragmenting in the shocked layer produced by the head-on collision.

Using $t_{\rm m}$ in equation (2.2), we define $\lq\lq$
the growth rate of the slab mass" as $\omega_{\rm m} \equiv 1/t_{\rm m}$.
Assuming $d \sigma / dt \simeq \sigma / t$, where $t$ denotes the time interval
measured from the onset of the collision, we have
$$\omega_{\rm m} = {1 \over t},
\eqno(5.2)$$
which monotonically decreases with time. We note, while, that the maximum
growth rate, $-i \omega_{\rm max}$,
of the instability increases with time, since $\sigma$ and $L$ increase
as the collision proceeds [See also e.g. figure 3.]. At the beginning of
the collision, $\omega_{\rm m} \gg - i \omega_{\rm max}$, and the slab mass
grows more rapidly than the perturbation does, and thus the slab is
practically stable.  When $-i \omega_{\rm max}$ exceeds $\omega_{\rm m}$,
the perturbation grows appreciably. Equating $\omega_{\rm m}$ with
$-i \omega_{\rm max}$, we meet the time $t_0$ when the perturbation starts
to grow,
$$t_0 = 1/ (- i \omega_{\rm max}).
\eqno(5.3)$$
We have then the Jeans mass $M_{\rm J}$ (5.1) in which
$\lambda_{x \, {\rm J}}$, $\lambda_{y \, {\rm J}}$, and $\sigma$ are evaluated
at $t=t_0$.

Using quantities defined in section 2, we have
$$ \sigma = {2 \rho_c c t \over M_{\perp}},
\eqno(5.4)$$
where we have used the relationships: $\sigma \simeq t d \sigma / dt$,
$d \sigma / dt \simeq 2 \rho_0 c M_{\perp}$,
$\rho_1 / \rho_0 \simeq {M_{\perp}}^2$, and $\rho_1 \simeq \rho_c$,
when $L \ll H_{\rm R}$.
Substituting $\sigma$ in equation (5.4) into (4.5), we have
$$-i \omega_{\rm max} = {2 \pi G \rho_c \over M_{\perp} \sqrt{1+ {M_{\parallel}
}^2}}t.
\eqno(5.5)$$
{}From equation (5.3) and (5.5), $t_0$ is written as
$$t_0 = {\left( M_{\perp} \over 2 \pi G \rho_c \right)}^{1/2}
{( 1 + {M_{\parallel}}^2)}^{1/4}.
\eqno(5.6)$$
Note that this is comparable to the free fall timescale,
when $M_{\perp} \simeq 1$.
Using the relations:
$k_{x \, {\rm max}} = \pi G \sigma
/c^2 (1+ {M_{\parallel}}^2)$ [from equation (4.6)], and
$k_{y \, {\rm max}} = \pi G \sigma /c^2$ [from equation (4.3)],
we have, respectively,
$$ \lambda_{x \, {\rm max}} = {c M_{\perp} \over G \rho_c t_0}
(1 + {M_{\parallel}}^2) =$$
$$\hbox{       } = 2 \pi
{H}{M_{\perp}}^{1/2}
{(1+ {M_{\parallel}}^2)}^{3/4},
\eqno(5.7)$$
and
$$ \lambda_{y \, {\rm max}} = {c M_{\perp} \over G \rho_c t_0}
 = $$
$$\hbox{      } = 2 \pi {H}{M_{\perp}}^{1/2}
{(1+ {M_{\parallel}}^2)}^{-1/4},
\eqno(5.8)$$
where we have used equation (3.14).
These are roughly comparable to or greater than $H$,
when $M_{\perp} \simeq 1$.
Finally, we obtain the Jeans mass from equation (5.1),
$$M_{\rm J} = 2 {\pi}^2 \rho_c {H}^3 {M_{\perp}}^{1/2}
{(1 + {M_{\parallel}}^2)}^{3/4},
\eqno(5.9)$$
where we have used
$$ \sigma (t = t_0) = 2 \rho_c {H}
{M_{\perp}}^{-1/2} {(1+{M_{\parallel}}^2)}^{1/4}.
\eqno(5.10)$$
Note that when $M_{\perp} \gg 1$, we have $L(t=t_0) \ll H \simeq H_{\rm R}$
from equation (5.10) and $\sigma \simeq 2 \rho_c L$.
When $M_{\parallel}$ tends to zero, equations (5.6) through (5.9) are
reduced to equation (1) in Whitworth et al.(1994),
except for numerical factors.

The Jeans mass increases as $U$ and $\Omega$ increase, and
therefore the more massive gaseous clump would be generated in the gaseous
slab at the interface of the obliquely colliding gas clouds.
When the collision is strong and the resultant gaseous slab is thin,
the perturbed motion normal to the shock front due to the Coriolis
force is suppressed and thus the Jeans mass tends to
that of the case of $U \not= 0$ and $\Omega = 0$ in subsection 4.2.
We have derived it as analytical form in equation (5.9).
In this case, as is shown in equations (5.7) and (5.8),
the clump formed by the gravitational instability is elongated
along the $x$-axis and rotates around the $y$-axis.

\section{Conclusion and Discussion}

We have worked on the gravitational instability of a gaseous slab with
shear and rotation whose axis lies within and parallel to it.
It is shown that more massive gaseous clumps are generated as the shear
becomes stronger. The shear may be understood as ensemble of rotating
eddies, contributing to increasing the effective sound speed. When the
gaseous slab is thin and pressed together with two plane-parallel
solid surfaces, the rotation affects little on the instability,
because the perturbing force or the Coriolis force due to the rotation
is directed only normal to them. We believe that the applied plane-parallel
solid surfaces are realistic enough to mimic the plane shock front which
is stable against its deformation. We briefly touch upon below,
what difference comes out from our results
when we take the free boundary condition where the pressure always
balance across the flexible and free surfaces.

\subsection{On the Boundary Condition Mimicking the Shock Front}

\begin{figure}
\vspace*{6cm}
\caption{The same figure as figure 4 but for the free boundary. The wavenumber
of this perturbation is now $k_x = 0.696$ (and $k_y = 0$).}
\end{figure}

In figures 2 and 3 we reproduce Elmegreen and Elemgreen's dispersion relation
(1978) in thin lines,
where $U = \Omega = 0$ and the free boundary condition are adopted.
The dispersion relations in figure 2 show that the difference between the
boundary conditions is more considerable as the slab thickness $L$ decreases:
compare (a,c) for $L=1.0$ and (b,d) for $L=0.5$. We find in figure 3 that
the maximum growth rate $-{\omega_{\rm max}}^2$ and their corresponding
wavenumber are independent of the boundary conditions as far as
$ L \, \ga \, 1 $.  They, however, depend strongly on the boundary
conditions when $ L \, \la \, 1 $.  When the boundary is \lq\lq free \rq\rq ,
$-{\omega_{\rm max}}^2$ does not change in the wider range of $L$,
while $k_{x \, {\rm max}}$ increases as $L$ decreases (Elmegreen and Elmegreen
1978; Lubow and Pringle
1993). As many authors suggested, these properties seems due to the instability
whose density remain nearly constant while their configuration tend to round
the structure (e.g. Vishniac 1983). Delineating the same figure as figure 4 but
now for the free boundary, we confirm this instability in figure 14.
The velocity field shows that the perturbed mass tend to round.

The fixed and free boundary conditions would be two extreme cases
representing a surface, since the former mimics completely rigid one,
and the latter flexible one. The real structure of the shock front
must be in between these two extremes (See Welter 1982).  We expect
that the real boundary is closer to the rigid one when
two clouds collide with high velocities and the shock wave is strong.

\subsection{Kelvin-Helmholtz Instability in the Slab
with Shear and No Rotation}

We have discussed exclusively the gravitational instability of the gaseous
slab with shear and rotation. However, the basic equations in section 3
can automatically deal with the Kelvin-Helmholtz (K-H) instability,
and it is appreciable when the wave length of the perturbation is shorter,
$k_x \geq 1$, and the velocity gradient of the shear, $d u_0 /dz$, is larger.
In order to study the Kelvin-Helmholtz instability we made a slab model
with strong velocity shear
in which the unperturbed shear flow is assumed to be
$$ u_{0} (z) = U \tanh \left( z \over l \right),
\eqno(6.1)$$
with $l \ll L$.

\begin{figure}
\vspace*{6cm}
\caption{Dispersion relations between $k_x$ and ${\rm Re} \, (-\omega)^2$.
In the segments delineated by thick solid lines, ${\rm Im} \,
\omega = 0$, while by thick dashed lines, ${\rm Im} \, \omega \not= 0$.
We calculate dispersion relations in a right region of the thin
line.}
\end{figure}

\begin{figure}
\vspace*{8cm}
\caption{The cross section of the shearing slab disturbed
by the growing sound waves. Iso-density contours are delineated.
$k_y, L, l, M_{\parallel},$ and $\Omega$ are the same as in figure 15.}
\end{figure}

Figure 15 shows the relationship between $k_x$ and ${\rm Re}\, (- {\omega}^2)$,
when $k_y = 0$, $L = 1.0$, $l = 1/30$, $M_{\parallel} = 2.0$, and $\Omega = 0$.
We note that the ordinate is not measured in $-{\omega}^2$ as figure 2, 5, 8,
and 11, but in ${\rm Re} \, ( -{\omega}^2)$,
since $-{\omega}^2$ becomes the complex number.
The solid lines denote
the modes having real $ \omega ^2 $ while the dashed lines denote the
mode having complex $ \omega ^2 $.  Thus each solid line is two-folded
while each dashed line is four folded.  In order to understand the
network we labelled a solid line with a single character and a dashed
line with two characters so that each symbol denotes a single continuous
line.  In the limit of $ U \rightarrow 0 $ the zigzag line
$ Ga - G - Gb - G - \cdot $ approaches the dispersion relation
of the gravitational instability.
The sequences, $ Ga - a - ab - a - \cdot $,
$ \cdot - b - Gb - b - ab - \cdot $,
and $ \cdot - c - ac - c - bc - c - \cdot $ are the essentially the sound
waves.  These mode interact each other and form mixed modes, e.g.,
$ Ga $, $ ab $, $ bc $, etc.  This mode coupling is common in the
dispersion relation of the K-H instability in the slab with
boundary surfaces (Glatzel 1987a, b).

Figure 16 shows the density perturbations for the mode $ b $ of
$ \lbrack k _x , \, {\rm Re} ( - \omega ^2 ) \rbrack \, = \, (3.2, \, 0.62) $,
the mode $ c $ of
$ \lbrack k _x , \, {\rm Re} ( - \omega ^2 ) \rbrack
\, = \, (5.0, \, 0.47 ) $,
and the mode $ bc $ of
$ \lbrack k _x , \, {\rm Re} ( - \omega ^2 ) \rbrack
\, = \, (4.1, \, -1.0 ) $.
We find that these have periodic structure along the $z$-direction with
``wavelength" $2 L / n$ : $n = 2$ for the mode $b$ and $n = 3$ for
the mode $c$.
The mixed mode $bc$ have characters of both the parent modes, and
is similar to the mode $ b $ in the region $ z \, > \, 0 $
and while it is similar to the mode $ c $ in the region $ z \, < \, 0 $.

Since the mixed mode has a complex frequency, they grow or decay
while oscillating in time, i.e., propagating with the phase speed
$ {\rm Re} \, \omega / k _x $ in the $ x $-direction.  The density perturbation
of the mixed mode has a smaller relative amplitude than
the velocity perturbation of the mixed mode,
$ \vert \rho _1 / \rho _0 \vert \, < \, \vert \mbox{\boldmath$u$}_1 / c \vert$.
Thus, the growth of the mixed mode leads mainly to the diffusion of
the velocity shear and does not likely to the formation of
gravitationally bound systems.

\subsection{Non-Linear Evolution of the Unstable Mass}

We have shown that the gravitationally unstable clump collapses while
rotating, when the slab has the shear or rotation. As many authors
have pointed out, such a clump fragments into two parts during the collapse,
and forms a binary system (or multiple system).
Fujimoto and Kumai (1994) applies the results described in section 4
for the formation of binary star-clusters observed in the Magellanic Clouds.

\vskip72pt

We thank Yasuki Kumai for valuable discussions in particular on the
formation of globular cluster formation models.  We also thank
Fumitaka Nakamura and Tomoaki Matsumoto for advice on
the numerical problems and providing us their stability analysis code,
from which our numerical code developed.  This work is financially
supported in part by the Grant-in-Aid for General Scientific Research
(04452013).

\vskip82pt


\bigskip

\bigskip

{\bf Appendix. Proof of} $\mbox{\boldmath$u_1 = 0$}$
{\bf at Critical Wave Number when}
$\mbox{\boldmath$M_{\parallel} = 0$}$

\bigskip

The present appendix is to show that $\mbox{\boldmath$u$}_1 =
\mbox{\boldmath$0$}$ at the critical
wave number $\mbox{\boldmath$k$}_{\rm cr}$,
i.e., when the perturbation is marginally stable ($\omega = 0$).
We write the perturbed equations (3.16) through (3.20)
in the non-dimensional form as in the text, and substitute equations (3.21)
and (3.22) into them where we take $\omega = 0$ and $u_0 = 0$.
We have then, writing $\mbox{\boldmath$k$}_{\rm cr}$ as
$\mbox{\boldmath$k$}$ for simplicity,

$$2 \Omega u_{z1} = i k_x \left( - \psi_1 -{ \rho_1 \over \rho_0} \right),
\eqno(A.1)$$

$$0 = i k_y \left( -\psi_1 - { \rho_1 \over \rho_0} \right),
\eqno(A.2)$$

$$-2 \Omega u_{x1} = {d \over dz} \left( - \psi_1 -{ \rho_1 \over \rho_0}
\right),
\eqno(A.3)$$

$$ i k_x \rho_0 u_{x1} + i k_y \rho_0 u_{y1} + {d \over dz} ( \rho_0 u_{z1})
= 0,
\eqno(A.4)$$

and

$$\left[ {d^2 \over dz^2} - ( {k_x}^2 + {k_y}^2 ) \right] \psi_1 = 2 \rho_1.
\eqno(A.5)$$

\bigskip

(1) ~When $k_y \not= 0$, we have from (A.2),

$$\rho_1 = -\rho_0 \psi_1.
\eqno(A.6)$$

Substituting equation (A.6) into (A.1) and (A.3), we have $u_{x1} = u_{z1}
= 0$ since $\Omega \not= 0$ is presumed. Using $u_{x1} = u_{z1} = 0$,
we have $u_{y1} = 0$, because of $k_y \not= 0$ in equation (A.4), and hence,

$$ \mbox{\boldmath$u$}_1 = 0.
\eqno(A.7)$$

\bigskip

(2) ~When $k_y = 0$ and $k_x \not= 0$, we combine equation (A.1) and (A.3) to
remove the term
$(-\psi_1 - \rho_1 / \rho_0)$,

$$ u_{x1} = { d \over dz} \left ( i u_{z1} \over k_x \right),
\eqno(A.8)$$

where we assume $\Omega \not= 0$ as before. Substituting equation (A.8) into
(A.4), we have $u_{z1} = 0$. Then, equation (A.1) gives the relation (A.6)
again. Substituting equation (A.6) into equations (A.1)
and (A.3), we have

$$u_{x1} = u_{z1} = 0.
\eqno(A.9)$$

Since we can choose $u_{y 1} = 0$ without loss of generality,
we have again equation (A.7).

\vfill

\eject

\end{document}